\documentclass[12pt,a4paper]{article} %klut12,reqno
 \usepackage{amsmath}
\usepackage{amssymb}
\usepackage{latexsym}
\begin{document}
\newcommand{\BEA}{\begin{eqnarray}}
\newcommand{\EEA}{\end{eqnarray}}
\newcommand{\BEAast}{\begin{eqnarray*}}
\newcommand{\EEAast}{\end{eqnarray*}}
\newcommand{\Bb}{\mathbb{R}}
\newcommand{\Bii}{\mathbb{Z}}
\renewcommand{\theequation}{\thesection.\arabic{equation}}
\newenvironment{Proof}{\par\noindent {\em Proof } \rm}{\par\medskip}
\newtheorem{Prop}{Proposition}
\newtheorem{Lem}{Lemma}
\newcounter{Actr}
\setcounter{Actr}{0}
\renewcommand{\theActr}{A\arabic{Actr}}
\newcommand{\calA}{{\mathcal A}}
\newcommand{\calO}{{\mathcal O}}
\newcommand{\calH}{{\mathcal H}}
\newcommand{\calF}{{\mathcal F}}
\newcommand{\calP}{{\mathcal P}}
\newcommand{\calK}{{\mathcal K}}
\newcommand{\calS}{{\mathcal S}}
\newcommand{\calKtild}{\widetilde{{\mathcal K}}}
\newcommand{\Dhat}{{\hat{\Delta} }}
\newcommand{\supp}{{\rm supp}}
\newcommand{\Hyp}{H_m^+}
\newcommand{\Fi}{\varphi}
\newcommand{\Ptild}{\tilde{P}_3^{\uparrow}}
\newcommand{\Itild}{\tilde{I}}
\newcommand{\Jtild}{\tilde{J}}
\newcommand{\spec}{{\rm spec}}
\newcommand{\spP}{{\rm sp}_P }
\title{No-Go Theorem for ``Free'' Relativistic Anyons in d=2+1}
\author{ Jens Mund \thanks{Supported by DFG, Sfb 288 
 ``Differentialgeometrie und Quantenphysik'' and by Studienstiftung des 
 Deutschen Volkes} \\
  Institut f\"ur theoretische Physik, \\ Freie Universit\"at Berlin, 
Arnimallee 14, 14195 Berlin, Germany\\
 e-mail: mund@physik.fu-berlin.de  }
\maketitle
\begin{abstract}
We show that a quantum field theoretic model of anyons cannot be ``free'' in 
the (restrictive) sense that the basic fields create only one-particle states out
of the vacuum. % (holomorphic functions)
\end{abstract}
\section{Introduction} \label{sec0}
In $2+1$ dimensional spacetime, the Bose-Fermi alternative does not exhaust 
all possibilities for particle statistics. Rather, statistics may be 
described by a representation of the braid group. This has first been realized 
by Leinaas and Myrheim~\cite{LM}, and quantum mechanical models with an 
abelian representation of the braid group have first been discussed 
by F.~Wilczek~\cite{W}, who coined the name {\em anyons} for such particles. 
In the framework of algebraic quantum field theory, Buchholz and Fredenhagen 
have shown~\cite{BF} that massive particle states might be localizable in 
spacelike cones 
%\footnote{
%A spacelike cone is a cone in Minkowski space which extends to infinity only 
%in spacelike directions. } 
only (rather than in bounded regions), which in $d=2+1$ allows for the 
possibility of braid group statistics~\cite{F,FM1}. 
The Hilbert spaces of scattering states for such theories are well known: 
they have been 
constructed by K.~Fredenhagen et al.~\cite{FGR} and for the abelian case also 
by Fr\"ohlich and Marchetti~\cite{FM2}, and coincide with the ones proposed 
by R.~Schrader and the author~\cite{MS}. In particular, the one-particle 
space $\calH^{(m,s)}$ is characterized by the mass $m$ 
of the particle and its spin $s$, which may take any real value. The special 
cases $s\in\Bii$ and $s\in\Bii+\frac{1}{2}$ correspond to bosons and fermions, 
respectively. For these cases the well-known free fields establish a 
``second quantization functor'', associating a quantum field to each particle 
type~$\calH^{(m,s)}.$ 

 For $s\not\in\frac{1}{2}\Bii$, in contrast, free fields with braid group 
statistics and satisfying the requirements from relativistic quantum field 
theory have not been constructed yet\footnote{
D.R.Grigore has constructed free fields in $d=2+1$ for any spin \cite{G}, but 
in contradiction to the generalized spin 
statistics connection holding in algebraic quantum field theory~\cite{F,FM1} 
they have bosonic statistics. Presumably, this is due to the fields having 
infinitely many components.}.  
The present  paper gives a reason for this: 
It is shown that, under very mild assumptions, no  free model with nontrivial 
abelian braid group statistics can exist.  Here, ``free'' is meant in the 
restrictive sense that the field algebra is generated by operators creating 
only one-particle states out of the vacuum. We recall that in the Bose and 
Fermi cases ($s\in\frac{1}{2}\Bii$) the Jost-Schroer theorem\footnote{
This theorem is due to B.~Schroer~\cite{S2} and
has been elaborated by R.~Jost~\cite{Jo} and further by 
K.~Pohlmeyer~\cite{Po}. For a didactic account, see \cite[Thm. 4-15]{SW}. 
O.~Steinmann has extended it to string-localized fields satisfying modified 
Wightman assumptions and Bose or Fermi statistics~\cite{St}.} asserts that 
this condition characterizes the free (Fock space) fields.  

One may still hope to find models of anyons which, though not satisfying the 
above criterion, are ``free'' in the less restrictive sense that their 
S-matrix leads to a trivial cross section.  
Indeed, an analogous situation is encountered in the context of integrable 
massive models of anyons in $d=1+1$, where one has models 
with a piecewise energy-independent S-matrix~\cite{S}, but none which are 
free in the strict sense --  even 
the models which are closest to being free show ``virtual particle creation''. 

Recent investigations on a localization concept based on the Tomita-Takesaki 
modular theory point into the same direction as the result of the present 
paper. This localization concept, described e.g. in 
\cite{BGL,S4}, equips each one-particle Hilbert space $\calH^{(m,s)}$ with a 
family of real subspaces 
$\calH_R^{(m,s)}(W)$ indexed by wedge regions $W$ in Minkowski space. 
 For $s\in\frac{1}{2}\Bii,$ this family can be extended to a {\em net} of real 
subspaces $\calH_R(\calO)$ indexed by the double cones $\calO$ in such a way 
that the free fields establish a functor from nets of real subspaces 
$\calO\mapsto\calH_R(\calO)$ to nets of von Neumann algebras $\calO\mapsto
\calF(\calO).$  For $s\not\in\frac{1}{2}\Bii$, on the other hand, the defect 
of $\calH_R(W_1\cap W_2)\subset \calH_R(W_1)\cap\calH_R(W_2)$ has been 
computed to be infinite (real) dimensional~\cite{M}. This implies that such a 
functor cannot exist for $s\not\in\frac{1}{2}\Bii.$ 

The article is organized as follows. 
Section~\ref{sec1} sets up our framework and formulates the assumptions. 
We first describe what is understood by relativistic anyons in  algebraic 
quantum field theory, and to which class of such models we restrict. Then 
these  assumptions are collected in (A0) to (\ref{cyclicity}). The field 
algebra is assumed to be generated by ``free fields'' (A6) localized in 
spacelike cones\footnote{
We speak of ``fields'' although these operators are {\em not} 
pointlike localized like Wightman distributions (not even stringlike, like in 
\cite{St}).}, whose asymptotic directions are equipped with ``winding 
numbers''. Our main technical assumption, reminiscent of the Wightman 
axioms, is that for two fields $\Fi_1,\,\Fi_2$ with spacelike separated 
localization regions, the norm of $\Fi_1\,U(x)\,\Fi_2\,\Omega$ 
is polynomially bounded in $x$ (\ref{Adist}). Here $U(x)$ represents the 
translation by $x\in\Bb^3$ and $\Omega$ denotes the vacuum vector.  

In section \ref{sec2} we arrive at the no-go result in two steps: 
If the asymptotic {\em directions} of the localization regions of two fields 
$\Fi_1$ and $\Fi_2$ are spacelike separated, their ``twisted commutator'' is a 
c-number function, even if the localization regions overlap 
(Proposition~\ref{PropJS}). 
This is completely analogous to the (first part of the) well known 
Jost-Schroer theorem. 
On the other hand, these commutation relations are consistent only in the 
case of permutation group statistics (Proposition~\ref{PropNoGo}). 

\section{Assumptions}
%Free Relativistic Anyons in $d=2+1$.} 
 \label{sec1}
In algebraic quantum field theory a model is specified by an observable 
algebra $\calA$ containing a family  $\calA(\calO)$  of a von Neumann 
algebras labelled by the open bounded regions $\calO$ in Minkowski space,  
which acts in a vacuum Hilbert space $\calH_0$  carrying a
representation of 
the Poincar\'e group, and which satisfies the Haag Kastler axioms~\cite{HK}. 
According to the Doplicher-Haag-Roberts theory, the 
set $\Delta$ of inequivalent irreducible representations of $\calA$
(superselection sectors) satisfying certain physical criteria, is 
in one-to-one correspondence with the set of inequivalent DHR endomorphisms of 
$\calA$, and the composition endowes $\Delta$ with the
structure of an abelian semigroup \cite{DHR,BF}. If it is actually a group, 
the model is called abelian. 
We will restrict to abelian models whose superselection sectors
are generated by exactly one {\em auto}morphism $\gamma$ of the observable 
algebra, so that in the present context $\Delta$ is assumed to be isomorphic 
to $\Bii$, or, if there is a natural number $N$ s.t. $\gamma^N\cong id$, to 
$\Bii_N$. Admitting the more general case of a finitely generated abelian 
group would not change the result of this article.  
In addition we assume $\gamma$ to be a covariant massive one particle
representation of $\calA.$ This means that the representation of $\calA$ 
corresponding to 
$\gamma$ intertwines the vacuum representation of the Poincar\'e group 
with a representation in which the energy-momentum spectrum contains an 
isolated mass shell as its lower boundary.  
The automorphism $\gamma^q$ inherits from $\gamma$ the property of being an 
irreducible covariant massive representation \cite{BF}. 
Buchholz and Fredenhagen have shown \cite{BF}, that such 
representations are localizable in spacelike cones, i.e. equivalent to
the vacuum representation  when restricted to the observable algebra of
the causal complement of any spacelike cone. 
This entails an  intrinsic notion of statistics associating  to each sector a  
representation of the  braid group, which for abelian models is one 
dimensional and characterized by the so called statistics phase 
$\exp(2\pi is)$. The values $s=0$ and $s=\frac{1}{2}$ correspond to bosons and 
fermions, respectively.  
In the case of a $\Bii_N$ superselection structure, i.e. if there 
is a natural number $N$ s.t. $\gamma^N\cong id$,  we further assume the 
statistics phase of $\gamma$ to be an $N^{th}$ root of unity. Thus we 
restrict to models not exhibiting the obstruction $\exp(2\pi isN)=-1$ 
analysed by K.-H.~Rehren in \cite[section 3.1]{Re}. 

A model of anyons satisfying the above requirements can equivalently be 
described (see, e.g. \cite[section 3.1]{Re}) by an algebra $\calF$ of 
unobservable charged field operators
transforming under the 
global gauge group~$\Dhat$ (the dual of the abelian group $\Delta$, i.e. in 
our case $U(1)$ or $\Bii_N$, respectively). Essentially, $\calF$ is the 
crossed product of $\calA$ with $\Delta.$ Within this setting, the assumptions 
made so far are made precise in the following conditions (A0) to 
(\ref{cyclicity}). 

\begin{list}{\bf (\theActr)}{}
\item {\em Framework.} \refstepcounter{Actr}
The Hilbert space $\calH$ of the model carries a representation of the global 
gauge group $\Dhat=U(1)$ or $\Bii_N $. The corresponding decomposition into 
charge superselection sectors is written as 
\begin{equation} \label{Hq}
\calH = \bigoplus_{q\in\Delta}\calH_q \;.\quad\mbox{ Here }\Delta=\Bii\mbox{ 
or }\Bii_N, \mbox{ respectively,}
\end{equation}
defined as the character group of $\Dhat$.   
$\calH$ also carries a unitary representation
$U$ of the universal covering group $\Ptild$ of the Poincar\'e group
s.t. the joint spectrum of the generators $P_\mu$ of the
translations $U(x)$ contains $\{0\}$ and an isolated mass shell 
$\Hyp:=\{p\in\Bb^3\,/\, p^2=m^2, p_0>0\}$, where $m>0$ is the particle mass:
\begin{equation} \label{spP}
\{0\}\cup\Hyp\,\subseteq \,\spec P \,\subseteq\; \{0\}\cup\Hyp\cup
\{p\in\Bb^3\,/\,p^2>M^2, p_0>0\}\;\mbox{ for some } M>m .
\end{equation}
The eigenspace to $0$ is one dimensional and is spanned by the vacuum vector
$\Omega\in\calH_{q=0}$. The subspace of $\calH$ belonging to the $\Hyp$ part
of the spectrum will be referred to as the one particle Hilbert space 
$\calH^{(1)}$ and is assumed to contain only states with charge $q=1$ and $-1$ 
(particles and antiparticles):
\begin{equation} \label{H1}
\calH^{(1)} =\{\phi\in\calH\,/\,P^2\phi=m^2\phi\,\}\;\subset\;\calH_1\oplus
 \calH_{-1}\; .
\end{equation}
%A $2\pi$ rotation is represented in $\calH$ by 
%$U(2\pi)=\sum_{q\in\Delta}e^{2\pi isq^2}\,P_{\calH_q}\,.$
\end{list}
Charge carrying anyonic fields are not localizable in compact spacetime 
regions. Rather, the localization of a field operator is characterized by a 
path in the set $\calK$ of spacelike cones and their causal complements. 
We denote by $\calKtild$ the set of homotopy classes of such paths. 
This concept is described in detail in \cite[see equ.(2.23)]{FGR}; here we 
will only have to compare paths $\Itild,\Jtild\in \calKtild$ ending at regions 
$I,J\in\calK$ with  either 
$I\subset J$ or $I\subset J'$ (the spacelike complement of $J$). Then the 
relevant information of the paths is just the ``cumulated'' angle, which can 
be unambigously compared in these cases. In this sense we understand the 
relations 
\begin{equation} \label{IinJ}
 \Itild\subset \Jtild\,, \quad\Itild=\Jtild+2\pi\,\quad\mbox{ and } \quad
\Itild< \Jtild\;.
\end{equation}
The field algebra is a net assigning to every $\Itild\in\calKtild$ a von 
Neumann
algebra $\calF(\Itild)$ of bounded operators in $\calH$ %the charged fields  
such that the following properties are satisfied:
\begin{list}{\bf (\theActr)}{}
\item {\em Isotony:} \label{isotony}  
$\calF(\Itild)\subset\calF(\Jtild)$ if $\Itild
\subset\Jtild$ in the sense of (\ref{IinJ}).  
\refstepcounter{Actr}
\item {\em Twisted  locality:} \label{locality}
$Z(\Itild,\Jtild)\,\calF(\Jtild)\,
 Z(\Itild,\Jtild)^{-1} \:\subset\:
\calF(\Itild)'\;$ if $J\subset I'$.\\ 
Here $\calF(\Itild)'$ denotes the commutant of $\calF(\Itild)$, and 
$Z(\Itild,\Jtild)$ is the ``twist operator''  defined (for the first time in 
  \cite{S3}) by 
\begin{equation}
 Z(\Itild,\Jtild)\,|\,\calH_q:= e^{-i\pi sq^2(2n+1)}\quad
   \mbox{ if } \Itild+2\pi n<\Jtild<\Itild+2\pi(n+1)  \label{Z} 
\end{equation}
\refstepcounter{Actr}
\item  \label{covariance} 
{\em Covariance under translations:} ${\rm Ad}U(x)\:\calF(\Itild)
 \:=\: \calF(\Itild+x)$ for all $x\in\Bb^3$. 
\refstepcounter{Actr}
\item  \label{inner}
{\em Internal symmetry:} $\calF(\Itild)$ is mapped onto itself 
under the action of the global gauge group $\Dhat$. 
\refstepcounter{Actr}
\item  \label{cyclicity}
{\em Reeh - Schlieder property:} The vacuum vector $\Omega$ is cyclic for each 
$\calF(\Itild)$. 
\refstepcounter{Actr}
\end{list}
The next assumptions express our definition of a free field and a temperedness 
condition. 
\begin{list}{\bf (\theActr)}{}
\item  \label{Afree} \refstepcounter{Actr}{\em Free field:}
Each $\calF(\Itild)$ is generated by a star stable set $\Phi(\Itild)$ of 
closed 
operators in $\calH$ (the free fields localized in $\Itild$)  
creating only one particle states out of the vacuum:
\begin{equation} \label{free}
\Fi\:\Omega\in\calH^{(1)}\quad\mbox{ for all }\Fi\in\Phi(\Itild)\;.
\end{equation}
\item  \label{Adist}\refstepcounter{Actr}{\em Temperedness condition:}
Let $\Fi_1\in\Phi(\Itild_1)$ and $\Fi_2\in\Phi(\Itild_2)$ with 
$I_1\subset I_2'$. 
Then $U(x)\,\Fi_2\,\Omega$ is in the domain of $\Fi_1$  for all 
$x\in\Bb^3$ and, furthermore, the function 
\begin{equation} \label{pol}
x\mapsto \left\|\,\Fi_1\,U(x)\,\Fi_2\,\Omega\,\right\|
\end{equation}
is locally integrable and polynomially bounded for large $x$. 
(Note that this can be violated only if the fields are
unbounded operators.) 
\end{list}
Properties (\ref{inner}) and (\ref{cyclicity}) imply a grading of the algebras 
$\calF(\Itild)$: every
$F\in \calF(\Itild)$ has a unique decomposition 
\begin{equation} \label{Fq}
F=\sum_{q\in\Delta}F_q \mbox{ , where }F_q \mbox{ carries charge }q\,,
\end{equation} 
i.e. $F_q:\calH_{q'}\to\calH_{q'+q}$ for all $q'\in\Dhat$. 
The same holds for the generating fields. 
In terms of this grading, twisted locality (\ref{locality}) entails the more 
familiar commutation relations for spacelike separated fields: 
Let $\Fi_1\in\Phi(\Itild_1)$ and $\Fi_2\in\Phi(\Itild_2)$ carry 
charges $q_1$ and $q_2$, respectively. Let further $I_2\subset I_1'.$ 
Then
\begin{equation} 
\Fi_1\,\Fi_2\,\Omega\;=\;R(\Itild_1,\Itild_2)^{q_1\cdot q_2}\,\Fi_2\,\Fi_1  
\,\Omega\;. \label{CR}
\end{equation}
Here, $R(\Itild_1,\Itild_2)$ is defined by 
\begin{equation}  
R(\Itild_1,\Itild_2):= e^{-2\pi is(2n+1)}\quad\mbox{ if }
  \Itild_1+2\pi n<\Itild_2<\Itild_1+2\pi(n+1)  \label{R}
\end{equation}
To prove equation~(\ref{CR}), one approximates the positive part of the polar 
decomposition 
of $\Fi_1$ by operators in $\calF(\Itild_1)$ to show $\Fi_1\,Z(\Itild_1,
\Itild_2)\,\Fi_2\,\Omega=Z(\Itild_1,\Itild_2)\,\Fi_2\,Z(\Itild_1,\Itild_2)^
{-1}\,\Fi_1\,\Omega$ which implies (\ref{CR}). 
\section{Results}
%No-Go for Free Anyons} 
 \label{sec2} 
We have chosen (\ref{Adist}) so that the free fields can be decomposed into 
creation and annihilation parts at least on vectors of the form $\Fi\,\Omega$ 
(Lemma~\ref{Lem+-}), which is enough to prove the (first part of the) 
Jost-Schroer theorem: 
If in the situation of equation (\ref{CR}), we translate the
localization regions such that they are not spacelike seperated any
more, only a multiple of $\Omega$ is added to the r.h.s. of that 
equation (Proposition~\ref{PropJS}). 

 For $\Fi\in\Phi(\Itild),$ let $\Fi(x):=U(x)\,\Fi\,U(x)^{-1}$ be the translated 
field, which is 
affiliated to $\calF(\Itild+x),$\footnote{(and {\em not} localized at $x$)} 
and let $H_m^-:=-\Hyp$ be the negative mass shell. 
\begin{Lem} \label{Lem+-}
Let $\Fi_1$ and $\Fi_2$ be space like seperated fields as in (\ref{Adist}) 
above. Then the $\calH$-valued function $\Fi_1(x)\,\Fi_2(y)\,\Omega$ is a 
tempered distribution, whose Fourier transform has support contained in 
$\left(H_m^-\cup H_m^+\right)\,\times\,H_m^+$. Let $F^+$ and $F^-$ be 
defined by the corresponding decomposition, i.e. 
\begin{eqnarray}
\Fi_1(x)\,\Fi_2(y)\,\Omega&=&
      F^+(x,y)+F^-(x,y)\,,\quad \mbox{ with }\quad\supp\,\widetilde{F^\pm}
\subset     H_m^\pm\,\times\,H_m^+\,.\label{F+-}\\
\mbox{ Then }&&\nonumber\\
 F^-(x,y)&=&\langle\,\Omega\,,\,\Fi_1(x)\,\Fi_2(y)\,\Omega\, 
 \rangle \,\Omega, 
\quad\mbox{ and } \label{f-}\\
\spP F^+(x,y)&\subseteq & H_m^+\,+\,H_m^+\;. \label{f+} 
\EEA
Here $\spP\psi$ denotes the spectral support of $\psi\,$ 
w.r.t. the energy momentum operators\footnote{ i.e. the set of points 
$p\in\spec P$ such that for any neighbourhood $V$ of $p$,  the spectral
projector $E_V(P)$ does not map $\psi$ to zero.}. 
\end{Lem}

\begin{Proof}. 
Let $f,g\in\calS(\Bb^3)$, the space of Schwartz functions. 
Due to the temperedness condition (\ref{Adist}), 
Riesz' theorem asserts the existence of a unique vector $F(f,g)\in\calH$ 
satisfying 
\[
%\begin{equation} \label{Ffg}
\langle\,\phi\,,F(f,g)\,\rangle\;=\;
\int dxdy \,f(x)\,g(y)\, \langle\,\,\phi\,,\,
  \Fi_1(x)\,\Fi_2(y)\,\Omega\,\rangle\;\mbox{ for all }\phi\in\calH\,,
\]
%\end{equation}
and whose norm can be estimated by
%\begin{equation} \label{fiffig}
\[
 \|F(f,g)\|\leq\int
dxdy\,|f(x)g(y)|\,\|\Fi_1(x)\Fi_2(y)\Omega\|\,.
\]
%\end{equation} 
By (\ref{Adist}), this shows that the linear map 
$F\,:\,(f,g)\,\mapsto\;F(f,g)$ is continuous in both entries w.r.t. the 
usual locally convex topology on Schwartz space, i.e. $F$ is a vector valued 
tempered distribution. 

To prove the statement on the support of its Fourier transform, let $f_1$ be 
in $C_0^\infty(\Bb^3)$, i.e. a 
smooth function with compact support, and $g\in\calS(\Bb^3)$. For all $A\in
\calF(\Itild_1+\supp f_1)'$ one verifies that
\[ \langle\,A\,\Omega,\,F(f_1,g)\,\rangle\,=\, 
  \langle\,A\,\tilde{\bar{f_1}}(P)\,\Fi_1^*\,\Omega\,,\,\tilde{g}(P)\,\Fi_2 
\, \Omega\,\rangle\,.
\]
Since  $\Fi_2\,\Omega$ and $\Fi_1^*\,\Omega$ are in $\calH^{(1)}$, 
the scalar product vanishes if $\supp \tilde{g}\cap\Hyp=\emptyset$ , 
or if $f_1$ is of the form $f_1=(\Box +m^2)f$ for some function $f\in
C_0^\infty(\Bb^3)$. Taking into account the fact that $\calF(\Itild_1+\supp 
f_1)'\,\Omega$ is dense in $\calH$, we conclude that 
\BEAast
 F((\Box+m^2)f,g)&=&0 \mbox{ for all }f\in C_0^\infty(\Bb^3),g\in
 \calS(\Bb^3)\;,      \quad  \mbox{ and} \\
 F(f,g)&=&0 \mbox{ for all } f\in C_0^\infty(\Bb^3)\;\mbox{ and }g\in
\calS(\Bb^3) \mbox{ with }\supp \tilde{g}\cap\Hyp=\emptyset\,.
\EEAast
By continuity of $F$, these two properties extend to all $f\in\calS(\Bb^3)$. 
Now we can proceed as in the case of Wightman fields \cite{SW}:

The support of the Fourier transform of $F$ consists of two disjoint sets 
contained in $\Hyp\times\Hyp$ and $H_m^-\times\Hyp$, respectively, 
thus defining the decomposition (\ref{F+-}). To analyse the energy momentum 
supports, we extend $F$ via the 
Schwartz nuclear theorem to a continuous linear map from 
$\calS(\Bb^3\times\Bb^3)$ into $\calH$, and note that for all $f,g\in\calS
(\Bb^3)$ we have 
$
e^{ix\cdot P}\;F(f\otimes g)\;=\;\widetilde{F}(\,e^{ix\cdot(p_1+p_2)}\;
 \tilde{f}\otimes\tilde{g}\,)\,. 
$
By linearity and continuity this yields
\begin{equation} \label{hfg}
h(P)\,F(f\otimes g)\,=\,\widetilde{F}\left(h(p_1+p_2)\cdot\,
\tilde{f}\otimes\tilde{g}\right) \;\mbox{ for all }h\in\calS(\Bb^3). 
\end{equation}
 From this equation and from the support properties of $\tilde{F}$ we conclude 
that 
$h(P)\,F^\pm(f,g)\,=\,0$ if $\supp h\cap (H_m^\pm+\Hyp)=\emptyset$, 
respectively. This shows that $\spP\,F^\pm(f,g)\subset H_m^\pm+\Hyp$. Since 
$H_m^- + \Hyp$ intersects the energy momentum spectrum (\ref{spP})
only in $\{0\}$, the vector $F^-(f,g)$ must be a multiple of the vacuum vector 
$\Omega$, the factor being $\langle\Omega\,,\,F^-(f,g)\,\rangle\,=\,\langle
\Omega\,,\,F(f,g)\,\rangle$. This shows that $F^\pm(x,y)$ are, like 
$F(x,y)$, well defined as {\em functions}, and have the properties (\ref{f-}) 
and (\ref{f+}), \hfill q.e.d.
\end{Proof}

Now we are ready to establish an analogon to the (first part of the) 
Jost-Schroer theorem: If in the situation of equation~(\ref{CR}), we translate 
the localization regions such that they are not spacelike separated any
more, only a multiple of $\Omega$ is added to the r.h.s of the commutation 
relation~(\ref{CR}). More precisely:  
\begin{Prop} \label{PropJS}
Let $\Fi_1\in\Phi(\Itild_1)$ and $\Fi_2\in\Phi(\Itild_2)$ carry charges $q_1$ 
and $q_2$, respectively. Let further $I_1\subset I_2'$. 
Then the fields satisfy the commutation relations 
\begin{equation} \label{JS}
\Fi_1(x)\,\Fi_2\,\Omega-R(\Itild_1,\Itild_2)^{q_1\cdot q_2}
\,\Fi_2\,\Fi_1(x)\,\Omega\;=\; c_{\Fi_1,\Fi_2}(x)\,\Omega\quad\mbox{ for all }
 x\in\Bb^3\;. 
\end{equation} 
Here $c_{\Fi_1,\Fi_2}(x)$ is the scalar product of the vacuum with the l.h.s. 
of equation~(\ref{JS}).  
\end{Prop}
Note that these commutation relations extend from $\Omega$ to 
$\bigcap_{i=1,2}\calF(\Itild_i+x_i)'\,\Omega$, which is dense in $\calH$ if 
$\Itild_1$ and $\Itild_2$ have ``equal winding numbers'' as explained before 
equation (\ref{D}) below. The
\begin{Proof} of this proposition is a straightforward adaption of
the proof of theorem~4-15 in \cite{SW} to the present anyonic case:  
Let $F_{1,2}^+(x,y)$ be the component of $\Fi_1(x)\Fi_2(y)\,\Omega$ 
whose Fourier transform has support in $\Hyp+\Hyp$ according to 
Lemma~\ref{Lem+-}, and $F_{2,1}^+(x,y)$ that of $\Fi_2(x)\Fi_1(y)\,\Omega$. 
Let further 
$R:=R(\Itild_1,\Itild_2)^{q_1\cdot q_2}$ be as defined in equation (\ref{R}).  
Lemma~\ref{Lem+-} asserts that 
\begin{equation} \label{3}
\Fi_1(x)\Fi_2\,\Omega - R\,\Fi_2\Fi_1(x)\,\Omega = 
c_{\Fi_1,\Fi_2}(x)\,\Omega +F_{1,2}^+(x,0)-R\,F_{2,1}^+(0,x).
\end{equation}
We have to show that the last two terms add up to zero.  
 For all $\psi\in \calH,$ the distribution 
$F_\psi(x)\,=\,\left\langle\,\psi\,,\,
 F_{1,2}^+(x,0)-R \,F_{2,1}^+(0,x)\,\right\rangle  $ 
is the boundary value of an analytic function, since its Fourier transform 
has support in the cone $V_+$ according to Lemma~\ref{Lem+-}~\cite[Thm.~IX.16]
{RS}.
 Further, equations (\ref{3}) and  (\ref{CR}) imply that $F_\psi$ vanishes on 
the real open set of points satisfying $I_1+x\subset I_2'$. 
Due to the edge of the wedge theorem, this forces $\hat{F}_\psi$ to vanish 
identically as a distribution~\cite[Thm.~2-17]{SW}, and hence as a function. 
\hfill q.e.d.
\end{Proof}
 From this proposition we obtain our main result, the no-go theorem for free 
anyons: 
\begin{Prop} \label{PropNoGo}  
Assume, some of the commutator functions $c_{\Fi_1,\Fi_2}(x)$ appearing in 
equation (\ref{JS}) of Proposition~\ref{PropJS} do {\em not} vanish 
identically in $x$. Then the commutation relations (\ref{JS}) are consistent 
only if the statistics parameter $s$ is half integer, i.e. only in the case of 
permutation group statistics. 
\end{Prop} 
{\em Remark. } The additional assumption of the proposition does not seem to 
be a severe restriction: if it was violated, we only needed a criterion 
allowing us to deduce commutation relations for the field algebra elements 
from those of the fields (like an energy bound satisfied by the fields), to 
conclude that the local observable algebras are commutative -- in 
contradiction to our general framework (A0) to (\ref{cyclicity}). 

\begin{Proof}. 
We choose two spacelike seperated spacelike cones 
$I_1,I_2\in\calK$, two corresponding paths $\Itild_1,\Itild_2\in\calKtild$ s.t. 
$\Itild_1<\Itild_2<\Itild_1+2\pi,$ and two fields $\Fi_1
\in\Phi(\Itild_1),\,\Fi_2\in\Phi(\Itild_2)$ together with  a translation 
vector $x\in\Bb^3$ s.t. $c_{\Fi_1,\Fi_2}(x)\neq 0.$ This presupposes that the 
charges $q_1$ and $q_2$ of the fields add up to zero.  Next we pick a cone 
$I_3\in\calK$ spacelike\footnote{
It is a necessary condition for the proposition that this 
geometric situation can be achieved. This is not
the case, e.g. if the ``free fields'' are only localizable in wedge regions.}
 to $I_1+x$ and $I_2$
 and s.t. $I_1+x\cup I_2\cup I_3$ are 
contained in some  $I_x\in\calK.$ 
Now we choose $\Itild_3\in\calKtild$ ending at $I_3$ with $\Itild_1<\Itild_3<
\Itild_2$ (to be definite) and a  field $\Fi_3\in\Phi(\Itild_3).$ Due to the 
localization properties of $\Fi_1(x)$ and $\Fi_3$, the commutator 
$c_{\Fi_1,\Fi_3}(x)$ vanishes.  
%Now we choose a  
%with charge $q_3=-q_2$, which entails $\Fi_3\,\Omega\,\bot\,\Fi_2\,\Omega.$ 
We have chosen the localization regions so that there is a path $\Itild_x\in
\calKtild$ containing $\Itild_1+x,\Itild_2$ and $\Itild_3$ in the sense of 
(\ref{IinJ}). Using isotony (A1), this implies that the subspace 
\begin{equation} \label{D}
 D:=\calF(\Itild_1+x)'\,\Omega\cap\,\bigcap_{i=2,3}\calF(\Itild_i)'\,
\Omega
\end{equation}
contains $\calF(\Itild_x)'\,\Omega,$ which is dense in $\calH$ due to locality 
(\ref{locality}) and cyclicity (\ref{cyclicity}). Thus, $D$ is a dense 
subspace on which equation (\ref{JS}) holds. 
Now let $\psi\in D.$ Denoting $R_{ij}:=R(\Itild_i,\Itild_j)^{q_i
\cdot q_j}$ and $c_{12}:=c_{\Fi_1,\Fi_2}(x)$,  
we get from Proposition~\ref{PropJS} and from equation~(\ref{CR}) 
\begin{align}
&\langle\,\Fi_1(x)^*\,\psi\,,\,\Fi_2\,\Fi_3\,\Omega\,\rangle \:=\nonumber\\
&\qquad=\left\langle\,\psi\,,c_{12}\, \Fi_3 \,
\Omega \,  \right\rangle 
+  R_{12}R_{13}R_{23}\,\left\langle\,\Fi_3^*\,\psi\,,\Fi_2\,\Fi_1(x)\,\Omega\,
 \right\rangle   \label{1st}\\
&\qquad=\left\langle\,\psi\,,R_{23}R_{13}c_{12}\,
\Fi_3\,\Omega \right\rangle 
+  R_{12}R_{13}R_{23}\,\left\langle\,\Fi_3^*\,\psi\,,\Fi_2\,\Fi_1(x)\,\Omega\,
 \right\rangle   .  \label{2nd}
\end{align}
In equation (\ref{1st}) we have first commuted $\Fi_2^*$  with $\Fi_1(x)^*$ 
and then $\Fi_1(x)$ with $\Fi_3,$ and in (\ref{2nd}) first $\Fi_3$ with 
$\Fi_2,$ and then $\Fi_3^*$ with $\Fi_1(x)^*.$ Note that all of the twofold 
products 
$\Fi_2^*\,\Fi_1(x)^*$ etc. are well defined on $D.$ Since $D$ is dense, we 
conclude that 
\[
c_{12}\, \Fi_3\,\Omega = 
R_{23}R_{13}c_{12}\,\Fi_3\,\Omega \;.
\]%the two vectors in the right entries of the scalar products in 
%(\ref{1st}) and (\ref{2nd}) coincide. 
By assumption, $c_{12}\neq 0,$ so that $R_{23}\,
R_{13}$ $= 1 $ follows. On the other hand,  according to 
equation (\ref{R}) we compute $R_{13}=\exp(-2\pi isq_1q_3)$$=R_{23},$ which 
implies $\exp(-4\pi isq_1q_3)=1.$ This is only possible if $s\in\frac{1}{2}
\Bii,$ 
since the charges $q_1,q_3$ only take the values $\pm1$ according to 
(\ref{H1}), \hfill q.e.d.
\end{Proof}
\paragraph{Acknowledgements.} It is a pleasure to thank B.~Schroer, to whom 
the basic idea leading to this article is due, him and H.-W.~Wiesbrock for the 
discussion of the modular localization concept, and also R.~Schrader, 
B.~Schroer, H.-W.~Wiesbrock and M.~U.~Schmidt for numerous helpful and 
stimulating discussions. 

\end{document}